\documentclass[aps,prx,groupedaddress,reprint,showpacs,pdftex]{revtex4-1}
\usepackage{graphicx}
\usepackage{tabularx}
\usepackage{multirow}
\usepackage{longtable}
\usepackage{dcolumn}
\usepackage{bm}
\usepackage{amsmath}
\usepackage{amssymb}
\usepackage{txfonts}
\usepackage[version=3]{mhchem}
\usepackage{url}
\usepackage[usenames,dvipsnames]{color}
\definecolor{Gray}{gray}{0.0}
\definecolor{lightGray}{gray}{0.35}

\begin{document}
\title{
  High-pressure BaCN$_2$ phases explored by genetic algorithm
}
\author{
  Peng Song$^{1}$,
  Mari Kawaguch$^{1}$,
  Yuji Masubuchi$^{2}$,
  Kenji Oqmhula$^{1}$,
  Kousuke Nakano$^{1,3}$,
  Ryo Maezono$^{1}$,
  Kenta Hongo$^{4}$\\}

\affiliation{\\
  $^1$School of Information Science, JAIST,
  Asahidai 1-1, Nomi, Ishikawa 923-1292, Japan\\
  \\
  $^2$Faculty of Engineering, Hokkaido University,
  N13 W8, Kita-ku, Sapporo 060-8628, Japan\\
  \\
  $^3$International School for Advanced Studies (SISSA),
      Via Bonomea 265, 34136, Trieste, Italy\\
  \\
  $^4$Research Center for Advanced Computing
      Infrastructure, JAIST, Asahidai 1-1, Nomi,
      Ishikawa 923-1292, Japan\\
  \\
}

\vspace{10mm}

\date{\today}
\begin{abstract}
Polymers containing nitrogen 
have attracted much attention 
in connection with their application to 
high energy density materials (HEDMs), 
in which energy is inherent in the triple bond. 
It is an interesting question whether such polymerized 
phases appear in the high-pressure phase of metal carbodiimide 
$M$CN$_2$, of which synthesis have been reported in recent years, 
but few studies have investigated 
the crystal structure at high pressure.
We have adopted a structure search based on the 
genetic algorithm coupled with {\it ab initio} 
electronic structure calculations to investigate 
possible crystal structures that may appear 
in the high-pressure phase of \ce{BaCN2}. 
The structure search successfully reproduced 
the previously reported crystal structures 
in the lower pressure range. 
With confirmed reliability of its predictive 
ability, the genetic search further 
predicts a polymerized phase with $Ima2$ appearing 
at higher pressure above 42~GPa. 
The polymerized phase takes the structure of 
a linear network of \ce{CN3} planar triangular units. 
It is understood that the anion site units CN$_2$, 
which are close to each other under high pressure, 
form covalent bonds directly with each other 
and stabilize the phase. 
\end{abstract}
\maketitle

\section{Introduction}
\label{sec.intro}
Polymetric nitrogen appeared in high pressure range
is expeced for the application
to the high energy density materials (HEDMs)
storing energy in the triple bond. 
\ce{N2} molecules in the high pressure phase has hence 
been studied theoritically
\cite{1992MAI,2009PIC,2008CHE,2010WAN,2015SUN}
and experimentally
\cite{2004ERE,2014TOM}, 
but found to be not suitable for the practical application 
because the triple bond gets dissapeared during
the decompression process.
As such, metal nitrides in high pressure range, instead,
have attracted interests for this purpuse. 
For instance, an azide ion compound \ce{LiN3} with 
\ce{N=N=N-} chain is predicted to undergo the phase transtion
to N-zigzag-chain-structure as the pressure gets increased 
\cite{2013WAN}.
It is reported for the matallic nitrides that
they undergo the polymerization at relatively lower pressure,
and the polymerized structure is kept even during the
decompression
\cite{2018HUA,2021NIU}.
The phase is actually interesting
because the structure is stabilized by
the network formed only by the anion,
being a rare case among inorganic compounds.
\vspace{2mm}
Metal nidrides with polyatomic anion-sites
have also been studied,
such as those with carbodiimide ions~(\ce{N=C=N^{2-}})
\cite{1978DOW,1964ADA,2000BECa,1994BER,1931COC,1997RIE,
2005LIU,2009LIU,2010TAN,2018DOL,2003REC,2006NEU,
2017CORa,2017CORb,2013KAL,2011UNV,2018MAS}
or cyanamide ions~(\ce{N-C#N^{2-}})
\cite{2002LIUa,2000LIU,2000BECb,2001BAL}.
Their structure are originally stabilized by
the 3-dimensional network formed the metal cation bonding. 
It is then interesting to see whether these compounds undergo
the transition to the polymerized phase where
the structure is kept instead by the anion only 
to form a network, but this point is not well
investigaged so far.
Since in these compounds, the anion sites are
replaced from 'dots' into the 'extended arms' of
the polyatomic unit, it is expected to realize
more easier condition to form the anion network
leading to the polymerized phase. 
The 'arms' also provides more internal degrees
of freedom such as the orientation and the bending,
giving more complicated competition between
enthalpy and entropy.
It is hence expected to get rich variety of
phases and corresponding responses
for functional materials. 

\vspace{2mm}
Only a few studies have examined the crystal structure 
of $M$-NCN ($M$CN$_2$) at high pressure, including 
the examples of cyanamide compounds, \ce{HgCN2} and \ce{PbCN2}. 
\ce{HgCN2} starts to decompose at 1.9~GPa. 
A theoretical prediction based on 
the density functional theory (DFT) reports 
that the lattice parameter of 
\ce{PbCN2} changes nonlinearly at 5~GPa. 
\cite{2002LIUb,2018MOL}
For the alkali metal carbodiimides, 
\ce{SrCN2} and \ce{BaCN2}, 
their synthesis has been reported.
\cite{2010KRI,1994BER,2018MAS} 
The tetragonal structure phase of \ce{BaCN2} 
is particularly interesting as a phosphor, \ce{BaCN2}:Eu, 
showing red luminescence up to 5.34~GPa.
This tetragonal structure is retained 
even after depressurization and 
does not decompose in this pressure range.
\cite{2020MAS}
The tetragonal \ce{BaCN2} is found to be stable than 
the rhombohedral \ce{BaCN2} over the entire temperature range 
up to the pyrolysis. 
It has also been reported to be applied as a flux 
for oxynitrides because it melts at temperatures 
lower than those at which the structure pyrolyzes.
\cite{2019HOSa,2019HOSb}
The solid solution based on \ce{BaCN2} with 
Ba partially replaced by Sr, 
Ba$_{0.9}$Sr$_{0.1}$CN$_2$, 
was synthesized and reported as a 
orthorhombic crystal structure.
\cite{2021MAS}
In these compounds, we expect a variety of phase 
transitions originating from the complexity of 
the internal degrees of freedom of the anion units, 
as described above. 
Investigating structural transitions 
of these materials as the dependence on the pressure 
is an important issue to be clarified. 
In particular, if polymerized phases emerge 
in the high-pressure region, as mentioned 
at the beginning of this section, 
we might see potential applications in HEDMs. 
However, the search for new structure 
of carbodiimide compounds is still unexplored, 
and in particular, there are still no reports 
of high-pressure polymorphs.

\vspace{2mm}
In this study, we performed a 
structure search of the polymorphs of \ce{BaCN2} 
over the high-pressure range 
using the genetic algorithm to evolute 
crystal structures. 
\cite{2021PS_RMa,2021PS_RMb,2022PS_KHa,2022PS_RMa,2022PS_RMb,2022GA_RM}
The structure search is confirmed to 
reproduce the previously reported crystal structures 
such as the orthorhombic Ba$_{0.9}$Sr$_{0.1}$CN$_2$, 
tetragonal \ce{BaCN2}, rhombohedral \ce{BaCN2}, 
and the structure of \ce{SrCN2} reported at low temperature. 
Getting confirmed the predictive reliability, 
the search predicts new polymerized structure 
appearing at high pressures above 42~GPa, where 
the planar triangular units of \ce{CN3} 
are polymerized to form linear chains. 
Compared with the mono-anionic azide ion compounds
\cite{2018HUA,2017SW_TC}, 
the present compound with polyatomic anion 
indicates that higher pressures are required 
to obtain the polymerized phase. 
It implies that the useful properties 
found in the phases with lower pressure, 
such as pressure-dependent fluorescence, 
are kept up to higher pressure. 
The properties of the compound MCN$_{2}$ has 
been reported up to 10~GPa in experiments 
and 15~GPa in calculations. 
The present work is the first to predict 
what kind of structural phases may appear 
in the high-pressure range.

\section{Method}
The evolutionary algorithm (EA) implemented in Universal 
structure predictor: evolutionary Xtallography software (USPEX) 
was used to predict the ternary nitide BaCN$_{2}$ at the pressures 
of 5, 10, 25, 30, 40, 50 and 100 GPa.~\cite{2006GLA}
The first generation is generated 
at random from 500 structures, with each subsequent generation 
generating 50 structures with 30\% heredity, 50\% 
random, 10\% mutation, and 10\% soft mutation.
Each structure undergoes 4 steps of relaxation to 
minimize stress and force.

\vspace{2mm}
Energies in USPEX were evaluated uing
the 'Vienna Ab Initio Simulation Package' (VASP) code
~\cite{1993KRE,1994KRE,1996KRE_a,1996KRE_b}
at the level of density functional theory (DFT).
Our DFT simulations adopted 
the Perdew-Burke-Ernzerh (PBE)~\cite{1996PER} functional
within the generalized gradient approximation (GGA). 
For predicted structures, 
structural relaxations and electronic structures
were also performed at the same level of theory as the structural exploation.
For Ba element, 5$s$ and 5$p$ states are treated as valence state.
A plane wave basis with  cutoff energy of 600 eV and 
Monkhorst-Pack $k$-meshes of
$8 \times 8 \times 8$
were sufficient for
getting the total energies converged within
an accuracy of 1 meV/atom.
~\cite{1976MON}
Fitting the Birch-Murnaghan EOS and calculating the 
bulk modulus were done with the open source program pymatgen.
~\cite{1947BIR,2013ONG}
The phonon properties are calculated using the finite 
displacement method used implemented in Phonopy 
with a supercell of $2 \times 2 \times 2$ dimension.~\cite{2015TOG}
\begin{figure}[htb]
  \begin{center}
    \includegraphics[width=1.0\hsize]{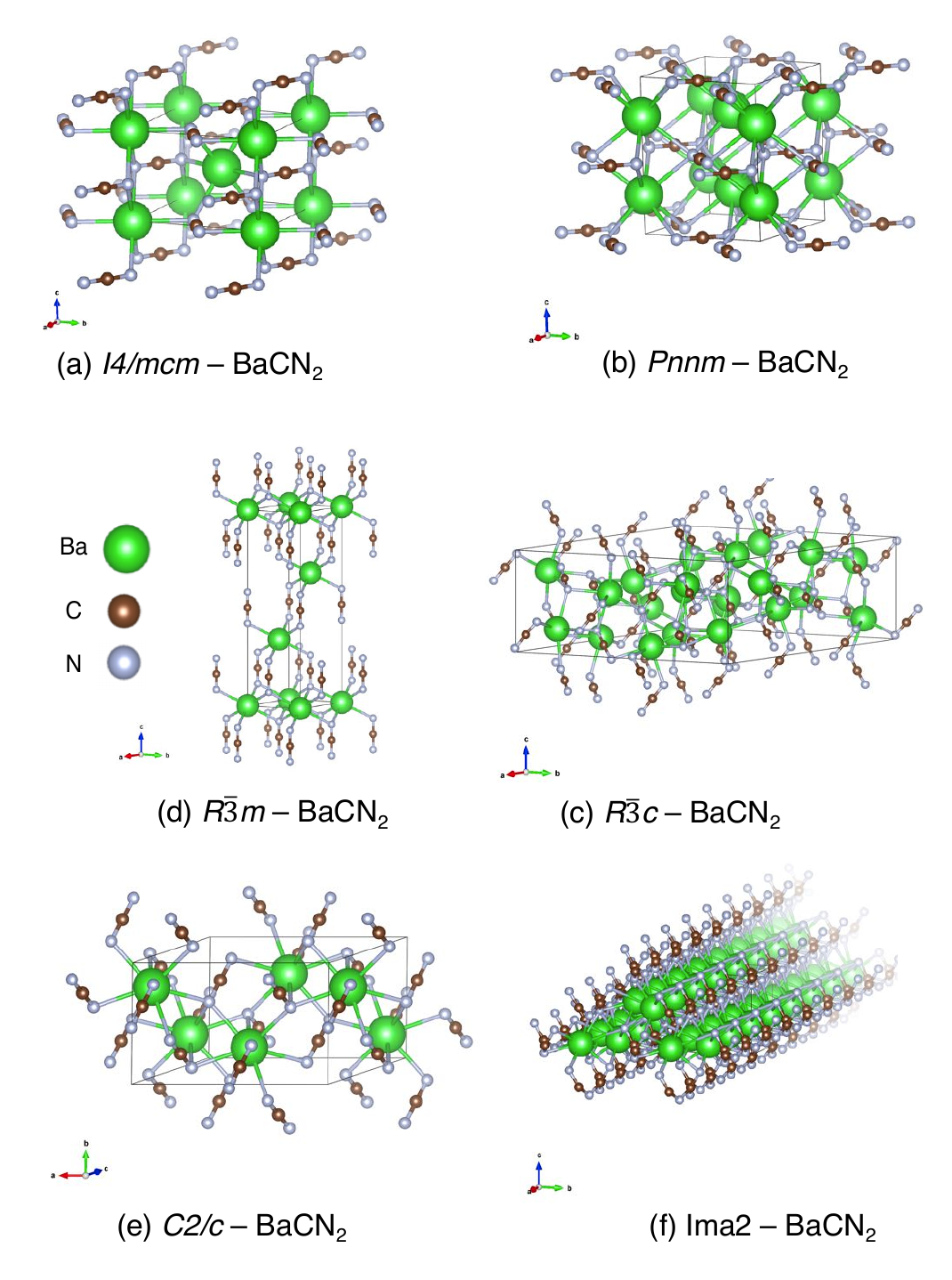}    
        \caption{
      BaCN$_{2}$ polymorphs 
      predicted by our genetic algorithm 
      at 0~GPa for BaCN$_2$, giving 
      (a)~$I4/mcm$ (140), 
      (b)~$Pnnm$ (58), (c)~$R\bar{3}c$ (167), 
      (d)~$R\bar{3}m$ (166), (e)~$C/2c$ (15), 
      and (f)~$Ima2$ (46) space groups.
    }
    \label{fig.structure}
  \end{center}
\end{figure}

\section{Results and discussion}
\label{sec.results}
\subsection{Predicted low-pressure phases}
We begin with the lower pressure range 
where several known structures are known 
to be compared with our predicted 
structures as shown in Fig.~\ref{fig.structure}. 
The figure shows a total of six polymorphs 
obtained by the genetic algorithm at 0~GPa for BaCN$_2$. 
Experimentally, tetragonal $I4/mcm$~(140) 
has been reported in the pressure region of $0\sim 6$GPa
\cite{2022MAsu}, which surely appears 
in the figure showing the consistency. 
The coordination number of Ba is eight,  
and it has a tetragonal inverted prismatic structural unit. 
Fig.~\ref{fig.structure}(b) is a rectangular $Pnnm$~(58) 
structure, which is found in Ba$_{0.9}$Sr$_{0.1}$CN$_2$ experimentally. 
In (c) and (d), rhombohedral 
$R\bar{3}m$ and $R\bar{3}c$ are found, 
which would correspond to the structure found in 
MgCN$_2$~($R\bar{3}m$), CaCN$_2$~($R\bar{3}m$), 
and BaCN$_2$~($R\bar{3}c$, experimentally. 
\cite{1994BER}

\vspace{2mm}
The structure, $C/2c$ shown as (e) has the same structural units 
as the tetragonal $I4/mcm$, but with a distorted structure 
with reduced symmetry. 
An interesting new structure is found as shown 
in Fig~\ref{fig.structure}(f), which is tetragonal $Ima2$. 
The structure has nine-coordinated Ba connected with N, 
forming linear chains of anions [CN$_3$]$_n$. 
\begin{figure*}
  \begin{center}
    \includegraphics[width=0.8\hsize]{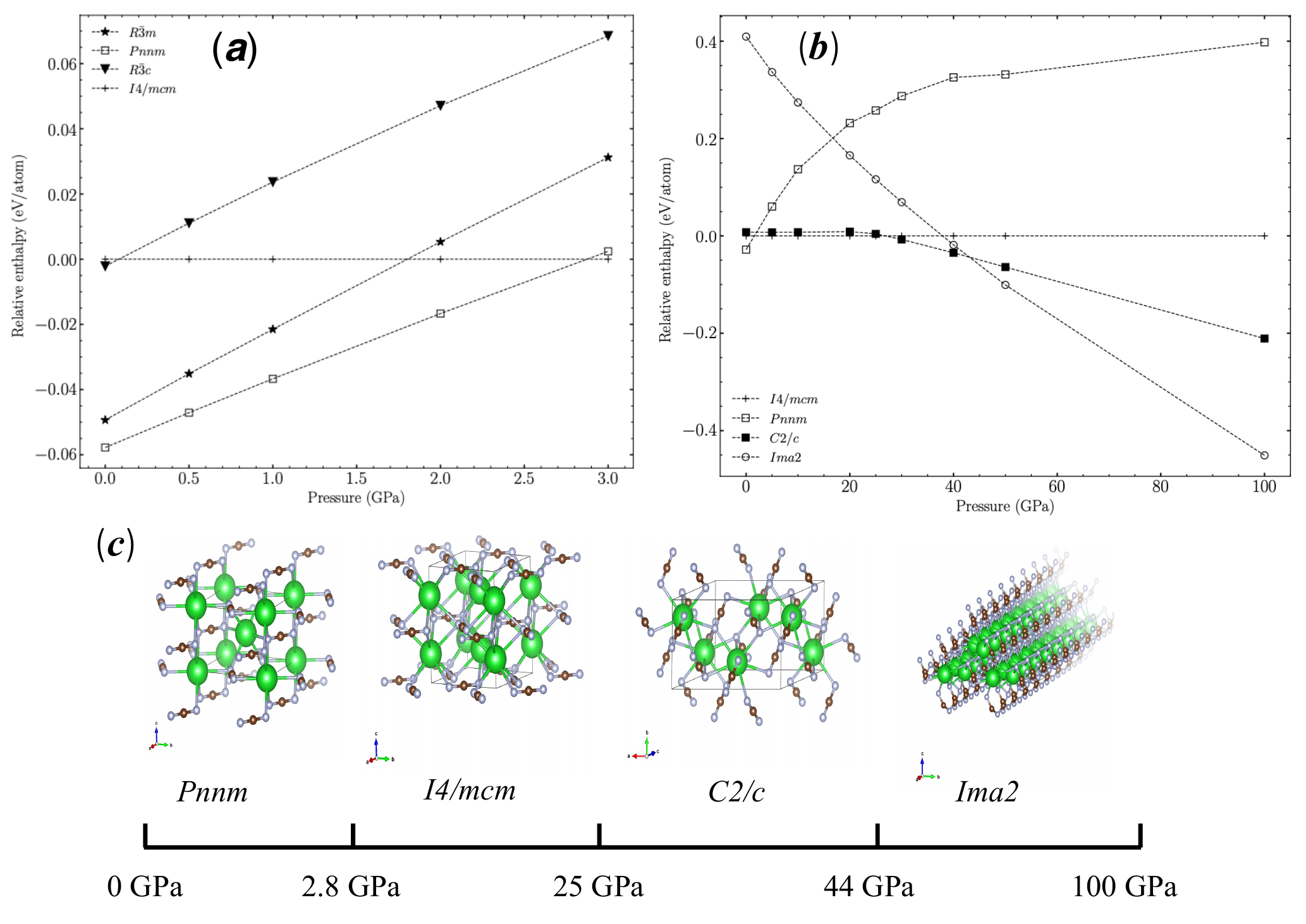}
    \caption{
      Relative enthalpies of BaCN$_{2}$ phases 
      compared with $I4/mcm$ as the reference zero, 
      plotted as a function of pressure [panel (a) and (b)]. 
      Panel (c) schematically shows the predicted 
      structural transitions as the pressure is applied.
    }
    \label{fig.enthalpy}
  \end{center}
\end{figure*}

\vspace{2mm}
To predict the structural transition 
depending on the applied pressure, 
we evaluated the enthalpy for each 
structure to be compared for 
thermodynamical stability
~\cite{2021PS_RMa,2022PS_KHa,2022PS_RMa}. 
The results are shown in Fig.~\ref{fig.enthalpy}. 
In the lower pressure region $0\sim2.8$GPa [panel (a)], 
$Pnnm$ is the most stable structure, and above that, 
$I4/mcm$ gets to be the most stable structure. 
Rhombohedral $R\bar{3}m$ and $R\bar{3}c$ 
structures are found not to be 
thermodynamically stable. 
The experimentally observed phases 
with these symmetry \cite{1994BER} 
are presumably realized as meta-stable 
structures. 
As the fact in experimental observations
~\cite{2022MAsu},
we confirmed $I4/mcm$ phase appears 
above 5~GPa with increasing pressure 
and kept even with decreasing 
puressure to 0~GPa as a hysteresis. 
At 0~GPa, the meta-stable $I4/mcm$ 
returns back to $Pnmm$ by re-baking 
at 450$^\circ$C 
in an argon atmosphere. 
The hysteresis would be allowed when 
the enthalpy difference between the 
competing phases is not so large, 
which can be discussed using 
the predictions in Fig.~\ref{fig.enthalpy} 
as given in the third paragraph of \S\ref{subsec.pressureDep}. 

\begin{figure*}
  \begin{center}
   \includegraphics[width=1.0\hsize]{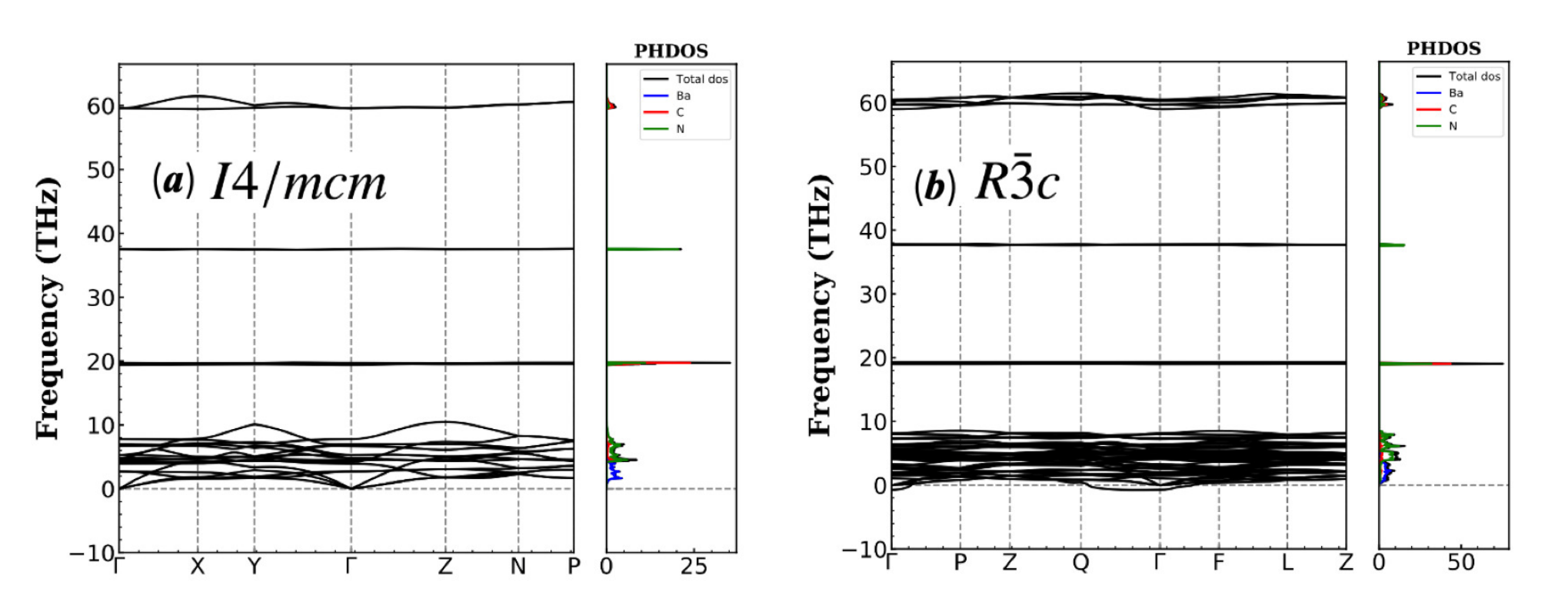}
    \caption{
      Phonon dispersion and phonon-DOS of (a)~$I4/mcm$ 
      and (b)~$R\bar{3}c$ at 0 GPa.
    }
    \label{fig.phonon}
  \end{center}
\end{figure*}

\vspace{2mm}
In Fig.~\ref{fig.phonon}, we examined the 
lattice dynamical stability for these 
candidate structures by {\it ab initio} 
phonon calculations. 
Panel~(a)~[(b)] shows the phonon dispersion 
of $I4/mcm$~[$R\bar{3}c$] structure at 0~GPa. 
We confirmed no imaginary mode appeared 
in $I4/mcm$ implying the dynamical stability, 
that is the case also for $Pnmn$ and 
$R\bar{3}m$ as shown in S.I. 
For $R\bar{3}c$, however, we observed some 
imaginary modes [panel (b)]. 
The imaginary modes for $R\bar{3}c$ have been 
confirmed to be kept even under increasing pressure. 
The present theoretical prediction is hence 
concluding that the $R\bar{3}c$ cannot exist 
as a stable phase. 
The phase is, however, observed as reported by 
single-crystal XRD experiments at ambient condition
\cite{1994BER}. 

\begin{figure}
  \begin{center}
    \includegraphics[width=1.0\hsize]{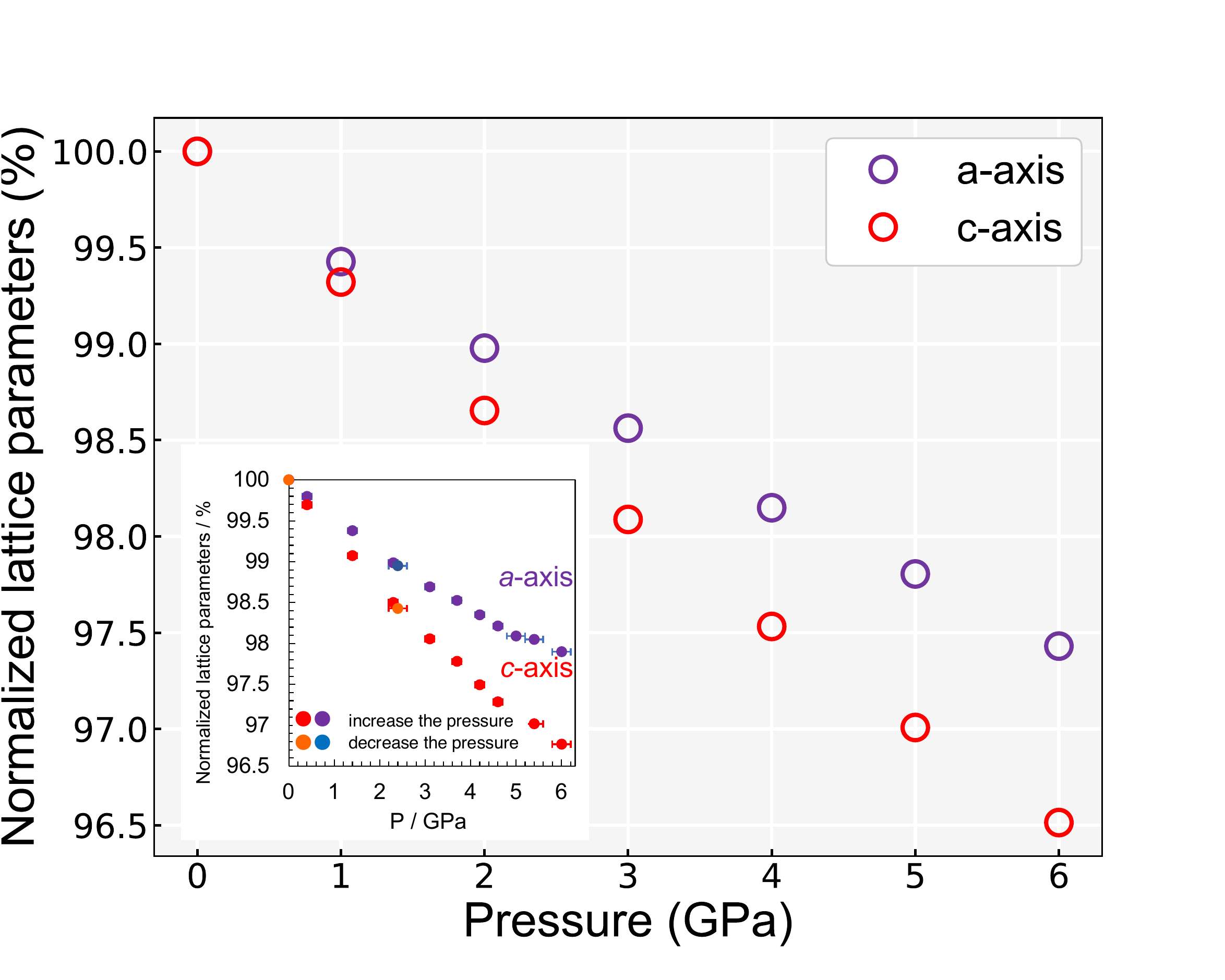}
    \caption{
Comparison of the pressure dependence of lattice constants 
between experiments and present calculations, 
evaluated for $I4/mcm$-BaCN$_{2}$. 
The vertical values are normalized by 
the reference values at 0~GPa. 
    }
    \label{fig.lattice}
  \end{center}
\end{figure}

\vspace{2mm}
Fig.~\ref{fig.lattice} shows the comparison 
of the pressure dependence of lattice constants 
between experiments and present calculations, 
evaluated for $I4/mcm$-BaCN$_{2}$. 
Fairly good coincidence to experiments 
would ensure the reliability of the present 
computational conditions with GGA-PBE 
exchange-correlation functionals. 
The dependence is summarized in terms of the 
bulk modulus being $69\pm 2$GPa~(experiments) 
to be compared with $71.8$~GPa~(calculations), 
being fairly good agreement as well (details are 
given S.I.). 

\subsection{Pressure-dependent stable phases}
\label{subsec.pressureDep}
The main interest of this study is the high-pressure 
phase of BaCN$_2$, which has not been studied experimentally. 
The pressure dependence of the relative enthalpies 
with that of $I4/mcm$ as the reference zero is shown 
in Fig.~\ref{fig.enthalpy}. 
Here, we exclude $R\bar{3}c$ and $R\bar{3}m$, 
which are already thermodynamically unstable 
than $I4/mcm$ in the low pressure region. 
As described in the previous subsection, 
$Pnnm$ is the most stable structure 
ranging in the low pressure region $0\sim 2.8$~GPa, 
and a phase transition to the tetragonal $I4/mcm$ is 
predicted at $2.8$~GPa. 
The tetragonal $I4/mcm$ is stable up to 25~GPa, 
where the phase transition to $C2/c$ occurs. 
Though the $C2/c$ has the same unit structure as $I4/mcm$, 
its symmetry drops from the four-fold ($I4/mcm$) 
to two-fold ($C2/c$) upon pressure application. 
The $C2/c$ is stable up to 42~GPa, 
above which the $Ima2$ phase appears 
at higher pressures. 

\vspace{2mm}
The newly predicted high-pressure phase $Ima2$ 
is a polymerized crystal formed by  
a linear chain binding anionic sites 
CN$_2$ only.  
The structure can be understood to emerged because 
of the high pressure making CN$_2$ come into close 
proximity and it is better to stabilize them 
by forming covalent bonds directly. 
This situation is similar for the high-pressure phase 
of CO$_2$, which has the same electron configuration as NCN$^{2-}$: 
It undergoes a phase transition from a molecular crystal 
to a covalent bond crystal at about 20~GPa 
\cite{1999SY_CM} 
when the pressure is increased because 
the closer distance between the elements 
make it better to form covalent bonds. 

\vspace{2mm}
In Sec.~Introduction, we raised the interest 
of the application viewpoint 
whether the polymerized phase realized 
by the high pressure can be quenched stably 
even with decompression, as realized in \ce{LiN3}. 
For this question, the behavior of enthalpy 
in Fig.~\ref{fig.enthalpy} gives negative support: 
Considering two different phases involved in hysteresis, 
enough small difference between each enthalpy
would allow the hysteresis with the compensation 
in entropy. 
As a matter of experimental fact\cite{2022MASu}, 
the hysteresis exists between $Pnnm$ and $I4/mcm$ 
in the low pressure region, where the enthalpy difference 
is $\sim$0.05~eV/atom. 
This magnitude would then be a reference to allow 
the hysteresis. 
Comaring with that, the enthalpy for the polymerized phase 
$Ima2$ increases too rapidly with decreasing pressure 
below below 20~GPa, giving much larger enthalpy difference, 
making it difficult to retain the hysteresis. 
Note that BaN$_6$=(Ba(N$_3$)$_2$) 
is another example of a high pressure phase 
showing the hysteresis to exist at ambient pressure 
(though it is not to be described as 
polymerized state but the compound of 
azide (NNN$^-$) ions). 
In this case, the energy difference between 
the phases involved in hysteresis is as small as 
0.033~eV/atom, and the structure of the high pressure phase 
is confirmed to be dynamically stable without imaginary 
modes in the phonon calculations, 
supporting the experimental fact 
that the high pressure phase is quenched. 

\vspace{2mm}
From the enthalpy comparison (Fig.~\ref{fig.enthalpy}), 
we have predicted the structural transitions 
$Pnnm\to I4/mcm\to C2/c\to Ima2$ as 
the applied pressure gets increased. 
The dynamical lattice stability of these structures 
was then verified by phonon calculations, 
confirming that no imaginary modes appeared 
for these structures supporting the phase stability (SI). 
The pressure-dependent phase diagram is summarized in 
Fig.~\ref{fig.enthalpy}. 
The coordination number increases as 
$Pnmm$~(6)$\to I4/mcm$~(8), $C/2c$~(8) $\to Ima2$~(9), 
being indeed consistent with the 
general statement that the coordination number 
increases as the applied pressure gets higher. 
The present prediction is a new finding because 
for the present compound, there has been 
a couple of reports upto 10~GPa~(experiment) 
and 15~GPa~(computational) so far.
\cite{2002LIUb,2018MOL}
Compared to the azide ion compounds 
with single atom anions
~\cite{2018HUA,2017SW_TC}
the present result predicts 
that higher pressure is required 
to obtain the polymerized phase 
when the anion site is replaced 
to the molecular unit. 
This can be understood as a result of 
increased internal degrees of freedom 
of the anion unit to accommodate 
more entropy. 
The structural transition requiring 
higher pressure implies that 
the pressure-dependent fluorescence properties 
as a useful application can robustly be maintained 
even under the higher pressure. 

\subsection{Electronic structures}
\begin{figure}
  \begin{center}
    \includegraphics[width=1.0\hsize]{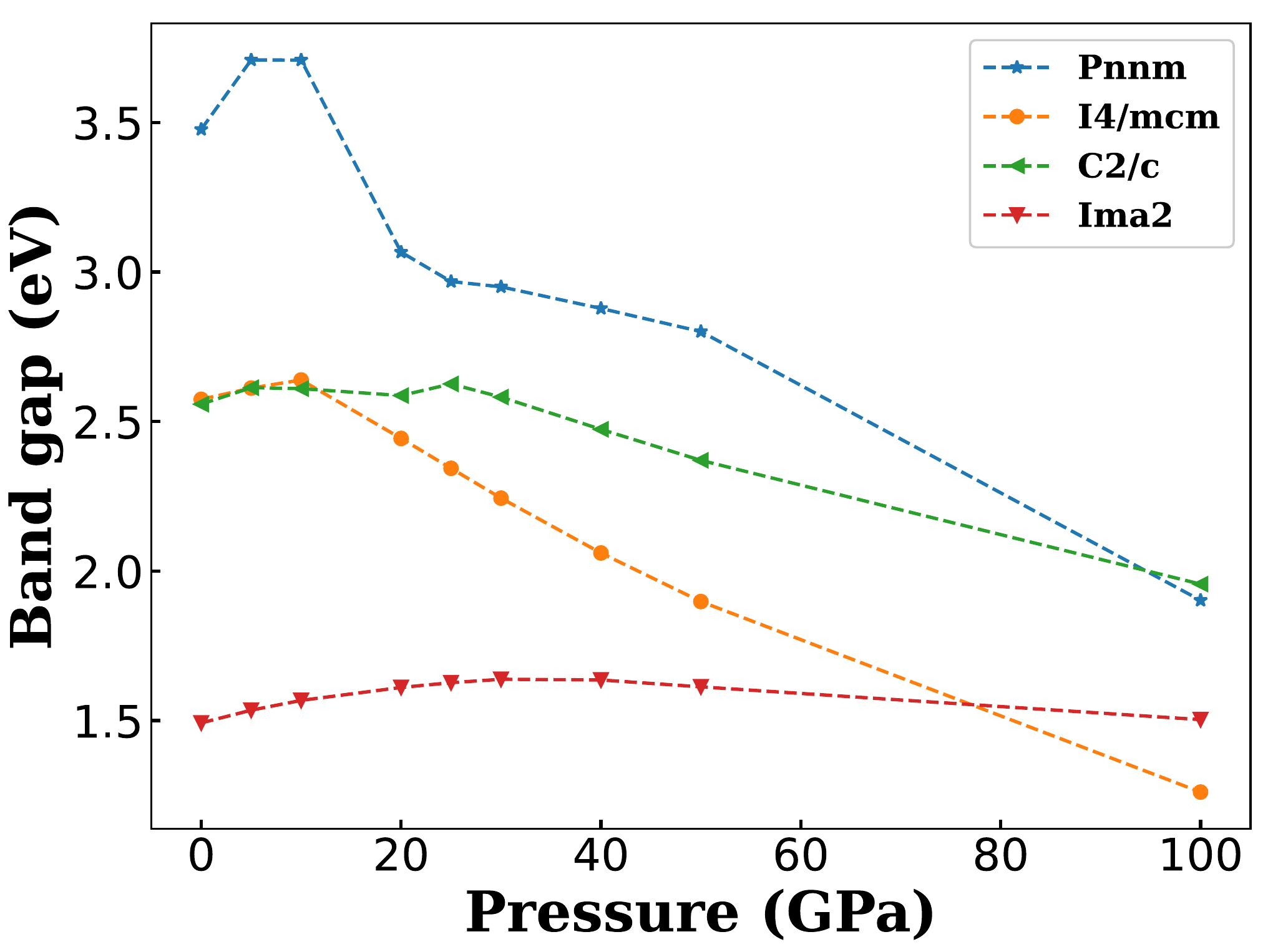}
    \caption{
    Estimated bandgap for each candidate structure 
    depending on the pressure. 
    }
    \label{fig.bandgap}
  \end{center}
\end{figure}

Fig.~\ref{fig.bandgap} shows the 
bandgap values predicted for each 
structure depending on applied pressure. 
The bandgap is a key quantity 
in the context of the application 
of the compounds as a phosphor~\cite{2020MAS}, 
where the transition of Eu$^{2+}$ (5$d\to$~4$f$) 
plays a role of its luminescence. 
For a phosphor, 
the energy change of the transition 
should be smaller than the bandgap 
of the matrix crystal, 
otherwise the conduction band overlaps  
the excited level of Eu$^{2+}$, leading to 
a non-radiative transition withiout 
any luminescence. 
It is, therefore, advantageous to 
have larger bandgap, and from this point of view, 
$Pnnm$ seems to be more promising 
as a phosphor than $I4/mcm$. 


\vspace{2mm}
It is worth noting that 
the behaviors in Fig.~\ref{fig.bandgap} 
for $I4/mcm$ and $C2/c$ are significantly different 
for these quite similar structures. 
The difference in the behavior would reflect 
how the overlap of orbitals get to be affected 
under the compression by the applied pressure. 
Compared to $C2/c$, $I4/mcm$ is then implied 
that its orbital overlapping has more 
sensitive structure to be affected by 
the compression. 
Looking at the polymerized structure $Ima2$, 
the behaviors in Fig.~\ref{fig.bandgap} 
is the least sensitive, which would be 
a consequence of its orbital overlapping 
being little to be affected. 
That is consistent with the linear behaviors in 
Fig.~\ref{fig.enthalpy}, from which we 
can conclude that the volume is almost 
unchanged and the overlapping situation 
is also the case. 

\vspace{2mm}
It is a general tendency for the bandgap 
to decrease as the applied pressure 
increases, being consistent with 
the behaviors in Fig.~\ref{fig.bandgap} 
for higher pressure range. 
In the lower pressure range, however, 
the bandgaps initially increase before 
decreasing in the further higher pressure range. 
This can be attributed to the anisotropy, 
without which we can simply understand 
that the isotropic compression leads to 
the monotonic reduction of the bandgap.
The initial uprising behavior is especially 
prominent for $Pnnm$ structure, for which
the {\it internal} anisotoropy
in the sense of orbital overlapping 
at the polyatomic anion unit would be
more significant than other structure. 


\section{Conclusion}
\label{sec.conc}
We investigated the high-pressure phase of metal 
carbodiimide {ce{BaCN2}} using genetic algorithm 
coupled with {\it ab initio} electronic structure 
calculations, with particular interest in whether 
the polymerized phase appears at high pressures. 
The structure search properly reproduced 
the previously reported crystal structures 
appearing in the lower pressure range, 
ensuring its reliability to some extent. 
The genetic structure search further 
predicts a polymerized phase with $Ima2$ 
appearing at higher pressure above 42~GPa. 
The polymerized phase takes the structure of 
a linear network of \ce{CN3} planar triangular units. 
It is understood that the anion site units CN$_2$, 
which are close to each other under high pressure, 
form covalent bonds directly with each other 
and stabilize the phase. 
The comparison of the behaviors of enthalpy 
as the pressure dependence 
for each structure phase implies that 
the polymerized phase is expected not to kept 
upon pressure reduction. 
Bandgaps were evaluated for each structure phase, 
showing that the phase realized in the lower pressure range 
has the largest bandgap, being advantageous for phosphor 
applications. 
  
\section{Acknowledgments}
The computations in this work have been performed
using the facilities of
Research Center for Advanced Computing
Infprastructure (RCACI) at JAIST.
K.N. acknowledges a support from the JSPS Overseas Research Fellowships, 
that from Grant-in-Aid for Early-Career Scientists (Grant Number JP21K17752), 
and that from Grant-in-Aid for Scientific Research(C) (Grant Number JP21K03400).
R.M. is grateful for financial supports from 
MEXT-KAKENHI (21K03400 and 19H04692), 
from the Air Force Office of Scientific Research 
(AFOSR-AOARD/FA2386-17-1-4049;FA2386-19-1-4015), 
and from JSPS Bilateral Joint Projects (with India DST). 
K.H. is grateful for financial support from 
MEXT-KAKENHI (JP16H06439, JP19K05029, JP19H05169, and JP21K03400),
and the Air Force Office of Scientific Research
(Award Numbers: FA2386-20-1-4036).

\bibliographystyle{apsrev4-1}
\bibliography{references}

\section{Supplemental Information}
\begin{table*}
 \begin{center}
   \caption{
     Crystal structures of
     BaCN$_{2}$ 
     predicted at each pressure~($P$).
     Lattice parameters ($a$, $b$ and $c$)
     are given in unit of $\AA$.
   }
     \label{table.crytal_struc}
\begin{tabular}{c|c|c|r|crrr}
  & &\multirow{2}{*}{$P$~(GPa)} & \multirow{2}{*}{Lattice parameters}
  & \multicolumn{4}{l}{Atomic coordinates (fractional)} 
\\ \cline{5-8}
&& &  & Atoms & $x$  & $y$  & $z$
  \\
\hline
BaCN$_{2}$    & $I4/mcm$ & 0 & \begin{tabular}[c]{@{}l@{}}
  $a = b = 6.07585$ \\
  $c = 7.30079$  \\
  $\alpha = \beta = \gamma =  90^{\circ}$\\
  \end{tabular}
  & \begin{tabular}[c]{@{}r@{}}
  Ba(4$a$)  \\
  C(4$d$)  \\
  N(8$h$)  \\
  \end{tabular}
  & \begin{tabular}[c]{@{}r@{}}
0.00000 \\
0.00000 \\
0.14452  \\
  \end{tabular}
  & \begin{tabular}[c]{@{}r@{}}
0.00000 \\
0.50000 \\
0.35548 \\
  \end{tabular}
  & \begin{tabular}[c]{@{}r@{}}
0.25000 \\
0.00000 \\
0.50000 \\
\end{tabular} \\
\hline
BaCN$_{2}$    & $Pnnm$ & 0 & \begin{tabular}[c]{@{}l@{}}
  $a =  5.58515 $ \\
  $b = 6.64885 $\\
  $c = 4.31738$  \\
  $\alpha = \beta = \gamma =  90^{\circ}$\\
  \end{tabular}
  & \begin{tabular}[c]{@{}r@{}}
  Ba(2$a$)  \\
  C(2$c$)  \\
  N(4$g$)  \\
  \end{tabular}
  & \begin{tabular}[c]{@{}r@{}}
0.00000 \\
0.00000 \\
0.18844  \\
  \end{tabular}
  & \begin{tabular}[c]{@{}r@{}}
0.00000 \\
0.50000 \\
0.59941 \\
  \end{tabular}
  & \begin{tabular}[c]{@{}r@{}}
0.00000 \\
0.00000 \\
0.00000 \\
\end{tabular} \\
\hline
BaCN$_{2}$    & $R\bar{3}c$ & 0 & \begin{tabular}[c]{@{}l@{}}
  $a = b =   15.16878 $ \\
  $c = 7.53574 $  \\
  $\alpha = \beta = 90^{\circ}$\\
  $ \gamma =  120^{\circ}$
  \end{tabular}
  & \begin{tabular}[c]{@{}r@{}}
  Ba(18$e$)  \\
  C(18$e$)  \\
  N(36$f$)  \\
  \end{tabular}
  & \begin{tabular}[c]{@{}r@{}}
0.00000 \\
0.00000 \\
0.04871  \\
  \end{tabular}
  & \begin{tabular}[c]{@{}r@{}}
0.20987\\
0.21645 \\
0.24013 \\
  \end{tabular}
  & \begin{tabular}[c]{@{}r@{}}
0.25000 \\
0.75000 \\
0.89123 \\
\end{tabular} \\
\hline
BaCN$_{2}$    & $R\bar{3}m$ & 0 & \begin{tabular}[c]{@{}l@{}}
  $a = b =   4.33547 $ \\
  $c = 15.47892 $  \\
  $\alpha = \beta = 90^{\circ}$\\
  $ \gamma =  120^{\circ}$
  \end{tabular}
  & \begin{tabular}[c]{@{}r@{}}
  Ba(3$a$)  \\
  C(3$b$)  \\
  N(6$c$)  \\
  \end{tabular}
  & \begin{tabular}[c]{@{}r@{}}
0.00000 \\
0.00000 \\
0.00000 \\
  \end{tabular}
  & \begin{tabular}[c]{@{}r@{}}
0.00000\\
0.00000 \\
0.00000 \\
  \end{tabular}
  & \begin{tabular}[c]{@{}r@{}}
0.00000 \\
0.50000 \\
0.41979\\
\end{tabular} \\
\hline
BaCN$_{2}$    & $C2/c$ & 25 & \begin{tabular}[c]{@{}l@{}}
  $a =  8.50422 $ \\
  $b =  5.28698$ \\
  $c = 5.82899 $  \\
  $\alpha = \gamma = 90^{\circ}$\\
  $ \beta =  128.438^{\circ}$
  \end{tabular}
  & \begin{tabular}[c]{@{}r@{}}
  Ba(4$e$)  \\
  C(4$d$)  \\
  N(8$f$)  \\
  \end{tabular}
  & \begin{tabular}[c]{@{}r@{}}
0.00000 \\
0.25000 \\
0.20203\\
  \end{tabular}
  & \begin{tabular}[c]{@{}r@{}}
0.29276\\
0.25000 \\
0.11282 \\
  \end{tabular}
  & \begin{tabular}[c]{@{}r@{}}
0.75000 \\
0.50000 \\
0.29637\\
\end{tabular} \\
\hline

BaCN$_{2}$    & $Ima2$ & 50 & \begin{tabular}[c]{@{}l@{}}
  $a = 4.39979  $ \\
  $b =  6.79556$ \\
  $c = 5.54136 $  \\
  $\alpha = \gamma = \beta =  90^{\circ}$\\
  \end{tabular}
  & \begin{tabular}[c]{@{}r@{}}
  Ba(4$b$)  \\
  C(4$b$)  \\
  N(4$a$)  \\
  N(4$b$)
  \end{tabular}
  & \begin{tabular}[c]{@{}r@{}}
0.25000 \\
0.25000 \\
0.00000\\
0.25000
  \end{tabular}
  & \begin{tabular}[c]{@{}r@{}}
0.15835\\
0.55642 \\
0.00000 \\
0.65223
  \end{tabular}
  & \begin{tabular}[c]{@{}r@{}}
0.50064 \\
0.48557 \\
0.11315 \\
0.28574
\end{tabular} \\
\hline
\end{tabular}
 \end{center}
\end{table*}

\subsection{Equation of states and bulk modulus}
\label{computational}
\begin{figure*}
  \begin{center}
    \includegraphics[width=\linewidth]{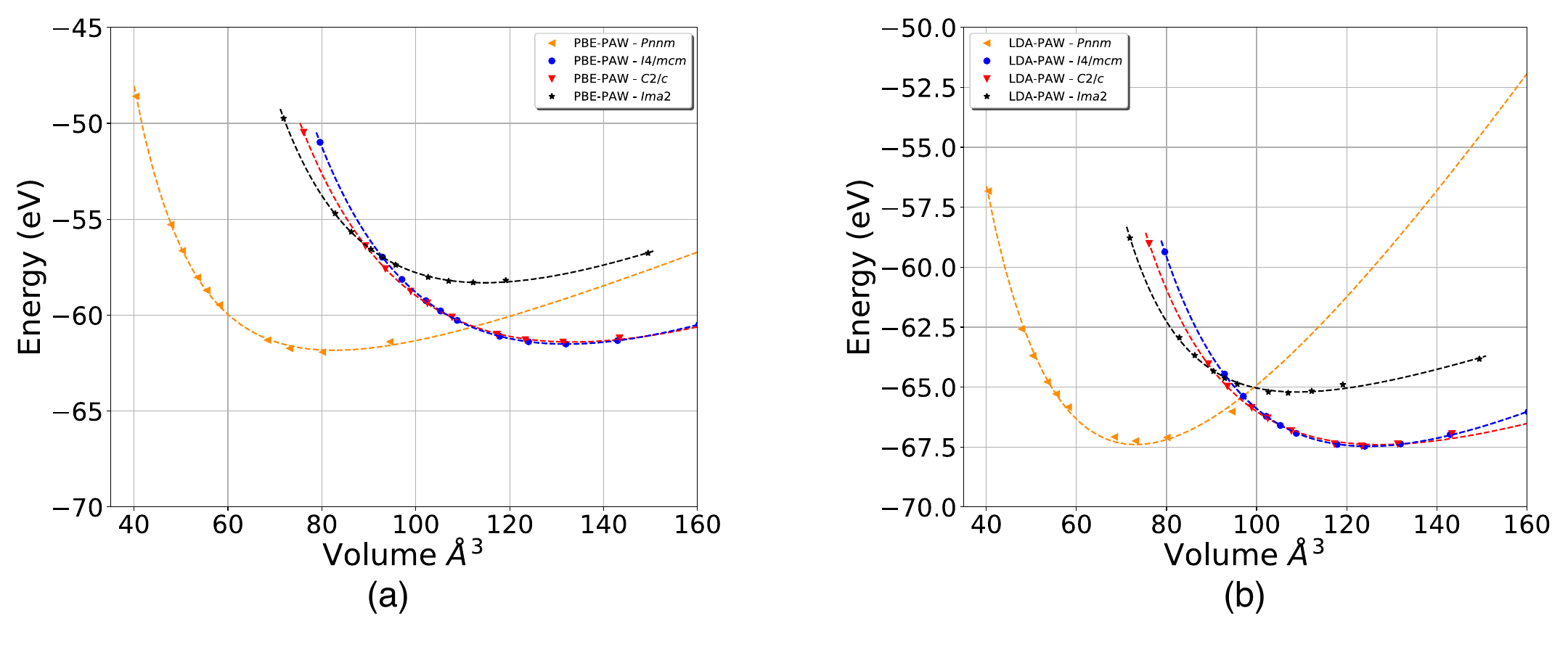}
    \caption{
EOS of BaCN$_{2}$ (a) GGA (b) LDA.
    }
    \label{fig.EOS}
  \end{center}
\end{figure*}

\end{document}